\documentclass[letterpaper,11pt]{article}

\usepackage{amsfonts}
\usepackage{amssymb}
\usepackage{graphicx}
\usepackage{eurosym}
\usepackage{float}
\usepackage{amsmath}
\usepackage[table]{xcolor}
\usepackage{tikz}
\usetikzlibrary{decorations.pathreplacing}
\usepackage{booktabs}
\usepackage{dcolumn}
\usepackage{rotating}
\usepackage{caption}
\usepackage{subcaption}
\usepackage[margin=1.25in]{geometry}
\DeclareMathOperator*{\argmin}{\arg\!\min}

\title{The Dirichlet Portfolio Model:\\Uncovering the Hidden Composition of\\Hedge Fund Investments}
\author{Laszlo F. Korsos\footnote{Korsos is at the Booth School of Business, 
University of Chicago, 5807 S. Woodlawn Avenue Chicago, IL 60637. Email: {\tt lkorsos@chicagobooth.edu}. I thank Jeffrey R. Russell, Nicholas G. Polson, Matt Taddy, Ralph S.J. Koijen, and Bryan T. Kelly for helpful comments, discussions, and suggestions.  I am grateful to Barclays Capital Indices for providing access to their fixed income index database.}}

\begin{document}
\maketitle

\begin{abstract}
\noindent Hedge funds have long been viewed as a veritable ``black box'' of investing since outsiders may never view the exact composition of portfolio holdings.  Therefore, the ability to estimate an informative set of asset weights is highly desirable for analysis.  We present a compositional state space model for estimation of an investment portfolio's unobserved asset allocation weightings on a set of candidate assets when the only observed information is the time series of portfolio returns and the candidate asset returns.  In this paper, we exhibit both sequential Monte Carlo numerical and conditionally Normal analytical approaches to solve for estimates of the unobserved asset weight time series.  This methodology is motivated by the estimation of monthly asset class weights on the aggregate hedge fund industry from 1996 to 2012.  Furthermore, we show how to implement the results as predictive investment weightings in order to construct hedge fund replicating portfolios.

\end{abstract}

\section{Introduction}
Over the past 20 years, the financial world has seen an enormous increase in demand for hedge fund products, thereby contributing to the estimated size of this global industry at approximately \$2.13 trillion as of April 1, 2012, according to Hedge Fund Research (HFRI).  These products intend to not only maximize returns on the assets under management during times of market boom, but also protect against losses during economic downturns.

High demand for access to these hedge fund products is manifested in the high value of fees charged to investors. On average, these fees come in the form of a 1-2\% management fee assessed on the total assets under management, in addition to a 15-25\% incentive fee on all capital gains.  Since it can be difficult for investors to assess the spectrum of individual hedge fund managers' skill, most investors tend to make their investments though a vehicle called a ``fund of funds''.  The purpose of these intermediaries is to evaluate individual hedge fund managers and then allocate an investors' assets across a broad spectrum of managers.  This division of investments is intended to diversify away risk associated with individual managers, and instead provide exposure to the returns of the hedge fund industry as a whole.  For these fund of funds services, a median 1.5\% management fee plus a 10\% capital gains fee is charged on top of the existing individual managers' fees.  These fees quickly add up and can easily eat away at any real profits that arise from capital appreciation on the invested funds.  Due to the combination of these high fee schedule layers, it is desirable to decompose and analyze the investment portfolios of these funds to determine if they are truly adding value for investors, or if similar strategies can be constructed with a much lower cost of investment.

As well, investors' investment goals and tolerance for risk exposures may not align with the incentive structure of hedge fund and fund-of-fund managers.  Therefore, a decomposition of a hedge fund's exposures to the risks arising from various asset classes is desired.  Of particular interest is how hedge fund managers respond to various macroeconomic events.  Using the model and estimation methodology presented here, we obtain a decomposition of the hedge fund industry's asset class risk exposures, which provide insight into their asset allocation process.  Interestingly, we find large increases in exposure to municipal bonds during the Dot-com Bubble decline in 2000-2001 and the recent global financial crisis from 2007-2012.

Another important feature of the hedge fund investing world is that return performance is only reported on a monthly or even quarterly basis.  Therefore, we can only observe how the fund has performed at certain discrete dates.  Between those dates, we cannot observe the current state of an individual's invested capital.  An investor could have doubled their money, or even lost half of their wealth overnight, but they will not know until the next reporting period.  If an investor had access to the invested asset value weightings, then they could compute estimates for intraperiod return and volatility values.  These values can have very important implications for current consumption choices, as well as risk management decisions.

This directly leads to a number of questions: Can we estimate what hedge funds are invested in, as well as how that asset allocation changes over time?  Then, using these estimates of asset allocation, can we generate intra-reporting-period return and volatility estimates?  That is, can we estimate how hedge funds are performing on a daily or even second-by-second granularity?  Furthermore, can we replicate this hedge fund portfolio in order to produce a similar series of returns, but through investing in easily accessible assets?

This portfolio estimation setup suggests a state space estimation problem where the latent compositional weights are required to sum to 1.  Due to this restriction, we venture beyond the classic Kalman Filter solution to estimate the weights (Kalman, 1960).  The results from Chipman and Rao (1964) and Tintner (1952) on Constrained Least Squares (CLS) estimation allow for this restriction in static models.  However, the CLS model does not allow for a dynamic compositional weight process.  In response, Chia (1985) and Simon and Chia (2002) present a solution with this restriction for dynamic models.  However, we show that these techniques do not perform well for this application.  Other notable work in compositional time series models are presented in Grunwald, Raftery, and Guttorp (1993) and Cargnoni, M{\"u}ller, and West (1997).  These focus on multinomial observational models of pure proportions or multinomial counts.  This work is in the spirit of those results, although we focus on univariate observations arising from a transformation of the latent compositional values.  The form of the generative dynamic model is as follows:
$$
w_t\sim Dir\left(\alpha\frac{w_{t-1}\circ(1+r_{PA,t-1})}{\sum_{i=1}^nw_{t-1,i}(1+r_{PA,t-1,i})}\right)
$$
$$
r_{HF,t}\sim t(w_t'r_{PA,t},\sigma_\epsilon^2,\nu)
$$
Our approach is to use the particle filtering methodology of Gordon, Salmond, and Smith (1993) to numerically solve the estimation problem on the portfolio weights $w_t$.  Also, making use of the particle filtering methods allows us to venture outside the simple Gaussian observational error assumption, thereby giving more suitable estimation results.

The remainder of this paper is structured as follows: Section 2 describes the statistical setup and motivation of the basic dynamic model.  Section 3 presents the fully specified Dirichlet Portfolio Model (DPM), the Sequential Monte Carlo approach for solving it, as well as an analytically solvable conditionally Normal approximation.  Section 4 outlines previous approaches to solve this problem, their respective drawbacks, and proposed improvements.  Section 5 compares the DPM to the other approaches under simulated and model hedge fund trading environments.  Section 6 uses actual hedge fund return data to estimate latent investment weights.  Section 7 outlines how these results can be used to estimate intraperiod hedge fund return and volatility values, as well as construct hedge fund replicating portfolios.  Finally, section 9 concludes.

\section{The Tracking \& Filtering Problem}

\subsection{General Setup}
First, consider the problem of estimating the latent weights on individual asset classes.  That is, at each time period we desire to combine prior information on these weights with new information introduced through observed overall hedge fund performance given the contemporaneous performance on the asset classes of interest.  This leads to defining a dynamic state space model of the following form:
$$
r_{HF,t}=f(w_t,r_{PA,t})
$$
$$
w_t=g(\mathcal{F}_{t-1})
$$
where $\mathcal{F}_t$ is the filtering of all information known at time $t$.  Hence, this includes all previous hedge fund index returns $r_{HF}$, palette asset returns $r_{PA}$, and palette asset weights $w$ up to and including time $t$.  That is, $\mathcal{F}_t=\left\{r_{HF,1},\dots,r_{HF,t},r_{PA,1},\dots,r_{PA,t},w_1,\dots,w_t\right\}$.  Note that $w_t=(w_{t,1},w_{t,2},...,w_{t,n})'$ is a $n\times1$ vector of the weights on each asset at the beginning of time period $t$, $r_{PA,t}=(r_{PA,t,1},r_{PA,t,2},..,r_{PA,t,n})'$ is a $n\times1$ vector of the palette asset returns over time period $t$, and $r_{HF,t}$ is a scalar value of the return on the hedge fund index over the same time period $t$.  The chronology of the time period notation is illustrated below:

\begin{figure}[H]
\centering
\begin{tikzpicture}[xscale=1.2]
\tikzset{position label/.style={below=10pt,text height=2ex,text depth=1ex}}
\draw [thick] (0,0) -- (12,0);
\draw [thick] (0,-0.2) -- (0,0.2);
\draw [thick] (4,-0.2) -- (4,0.2);
\draw [thick] (8,-0.2) -- (8,0.2);
\draw [thick] (12,-0.2) -- (12,0.2);
\node [position label] (t1) at (0,0) {$w_{t-1}$};
\node [position label] (t2) at (4,0) {$w_{t}$};
\node [position label] (t3) at (8,0) {$w_{t+1}$};
\node [position label] (t4) at (12,0) {$w_{t+2}$};
\draw [thick,decoration={brace,amplitude=6pt,mirror},decorate] (t1.north) -- (t2.north) node[midway,below=5pt] {$r_{PA,t-1},r_{HF,t-1}$};
\draw [thick,decoration={brace,amplitude=6pt,mirror},decorate] (t2.north) -- (t3.north) node[midway,below=5pt] {$r_{PA,t},r_{HF,t}$};
\draw [thick,decoration={brace,amplitude=6pt,mirror},decorate] (t3.north) -- (t4.north) node[midway,below=5pt] {$r_{PA,t+1},r_{HF,t+1}$};
\node[above] at (2,0) {$t-1$};
\node[above] at (6,0) {$t$};
\node[above] at (10,0) {$t+1$};
\end{tikzpicture}
\caption{\textsl{\small{Timeline Notation Illustration}}}
\label{1_timeline}
\end{figure}
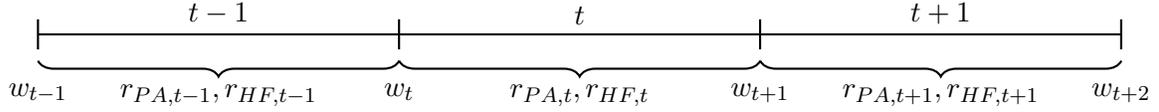

We can define a form for the observation function $f(\cdot)$.  Since the aggregate return on a portfolio of assets in simply the sum of the value weighted returns on the assets, the observation equation can be written as follows:
$$
r_{HF,t}=w_t'r_{PA,t}+\epsilon_t
$$
where $\epsilon$ is a stochastic term to be given a distributional form later.  This term is very important because it is unrealistic and potentially impossible to include all possible assets that a portfolio may be comprised of.  Therefore, it is necessary to allow for a term to pick up the variation in the observed index returns which is orthogonal to the palette asset returns.

Determining the form of the transition function, $g(\cdot)$, is a bit more challenging.  There may not necessarily be an exact science of how portfolio managers transition their asset weightings from period to period, but Amenc, Martellini, Meyfredi, and Ziemann (2010) suggest the following property:
$$
E[w_t|\mathcal{F}_{t-1}]=w_{t-1}
$$
This suggests that on average, portfolio managers keep the same asset value weightings from period to period.  This is a reasonable assumption, however it does introduce a subtle problem.  To illustrate this, take for example a portfolio of 2 assets where both are initially given equal value weighting (i.e. 50\% each).  Now, suppose that asset 1 yields a return of 0\%  and asset 2 yields a return of 100\% over a given time period.  Due to capital appreciation, assets 1 and 2 now have value weightings of 33.3\% and 66.7\%, respectively.  If the above property were employed in creating a transitional distribution, the prior expectation of the asset weights would both be 50\% (hence completely ignoring the idea of capital appreciation/depreciation).  This property hereby causes an artificial ``mean reversion'' effect on the asset weights since assets with relatively high return performance will be forced to have a relatively low prior in the next period, and vice versa.

Since an estimation procedure is desired which does not favor a ``mean reversion'' effect over a ``momentum'' effect, it is much more intuitive to implement a true random-walk process for the weights, which is what Amenc et al likely intended.  Since it is unlikely that the aggregate universe of portfolio managers consistently employs an asset allocation strategy which ignores capital appreciation, this is economically reasonable as well.  In order to account for capital appreciation and depreciation, the previous period's weight estimates, $w_{t-1}$ are updated by the relative increase in the observed period $t-1$ asset returns, $r_{PA,t-1}$.  This gives the following property:
\begin{equation}
E[w_t|\mathcal{F}_{t-1}]=\frac{w_{t-1}\circ(1+r_{PA,t-1})}{\sum_{i=1}^nw_{t-1,i}(1+r_{PA,t-1,i})}\label{1_rw}
\end{equation}
where $\circ$ is the Hadamard product.

It is important to note that the aggregate size of the hedge fund industry is about \$2 trillion.  Due to the very large nature of this aggregate hedge fund portfolio, it is unlikely that the entire industry could make major rebalancing shifts in the asset class weights from period to period.  That is, it would be very unlikely, and incredibly difficult for the entire industry to consistently employ either a ``mean reversion'' or ``momentum'' style strategy.  This is further, and potentially stronger support for the above property.

A stochastic component $\eta$ is incorporated into the weight transition in order to allow for the period-to-period uncertainty about transitional changes in the weights.  A distributional form will be imposed on this as well.

The general dynamic state space model for this problem is written:
\begin{equation}
r_{HF,t}=w_t'r_{PA,t}+\epsilon_t\label{1_obs}
\end{equation}
\begin{equation}
w_t=\frac{w_{t-1}\circ(1+r_{PA,t-1})}{\sum_{i=1}^nw_{t-1,i}(1+r_{PA,t-1,i})}+\eta_t\label{1_trans}
\end{equation}
$$
s.t.\quad\sum_{i=1}^nw_t=1
$$
Note that in this model there is no non-negativity constraint on the weights.  Negative values would imply a ``short'' weight on a palette asset.

\subsection{Estimation}

We are interested in solving for estimates of the asset weights conditional on all information available up to and including the current period.

First, because of the Markov property of the model, the true weights at time $t$ can be written as conditionally independent of all earlier times given information in the previous time $t-1$:
$$
p(w_t|w_0,...,w_{t-1},r_{PA,0},...,r_{PA,t-1})=p(w_t|w_{t-1},r_{PA,t-1})
$$
As well, the observation model at time $t$ is conditionally independent of all earlier times given information in the current time $t$:
$$
p(r_{HF,t}|w_0,...,w_{t},r_{PA,0},...,r_{PA,t})=p(r_{HF,t}|w_t,r_{PA,t})
$$
Therefore, the probability distribution over all states in the model is:
$$
p(w_0,...,w_t,r_{HF,1},...,r_{HF,t})=p(w_0)\prod_{\tau=1}^tp(r_{HF,\tau}|w_\tau,r_{PA,\tau})p(w_\tau|w_{\tau-1},r_{PA,\tau-1})
$$
In order to estimate the weights $w_t$ conditional on the information up to the current time $t$ we simply need to marginalize out the previous time periods.  Bayes rule gives the following expression:
$$
p(w_t|r_{HF,t},r_{PA,t})\propto\underbrace{p(r_{HF,t}|w_t,r_{PA,t})}_{Likelihood}\underbrace{p(w_t|r_{HF,t-1},r_{PA,t-1})}_{Prior}
$$
This is the ``update'' step, where the prior on the weights at the current time period is given by:
$$
p(w_t|r_{HF,t-1},r_{PA,t-1})=\int p(w_t|w_{t-1},r_{PA,t-1})p(w_{t-1}|\mathcal{F}_{t-1})dw_{t-1}
$$
The estimates of interest are obtained, $p(w_t|\mathcal{F}_t)$.

Under conditions of linearity and Normality this problem can be solved analytically with the Kalman filter.  However, if either of those conditions are violated, then the above densities are intractable and therefore approximate inference must be employed via Sequential Monte Carlo Methods.

\subsection{Prediction}

At a given point in time $t$, the posterior predictive distribution is used as our next period forecast.  This is given by:
$$
p(w_{t+1}|w_{t},r_{PA,t})
$$This is the same distribution as the prior for the next step ahead estimation problem.  This prediction problem is of special interest since access is not available to the aggregate of ``true'' hedge fund industry asset allocations, and therefore if we are to believe the assumption that the aggregate hedge fund industry does not (and possibly cannot, due to its large size) change asset class weightings very quickly, then the ``predictive'' return accuracy will give insight into how accurate the estimation technique is when using real world data.

As well, by constructing an estimation method for the relative portfolio weights, then these estimated latent weights can be used to project what funds may be invested in at any point in the future.  This gives the ability to estimate a distribution of potential latent intraperiod returns at any of these points:
$$
p(r_{HF,t}|r_{PA,t},\mathcal{F}_{t-1})=p(w_{t}|r_{PA,t-1},r_{HF,t-1})'r_{PA,t}
$$

\section{Sequential Monte Carlo Approach}
Herein, the proposed general dynamic model for this state space problem will have distributional forms imposed on the stochastic nature of the expressions in order to develop a feasible model for estimation.
\subsection{Model}
First, the weight transition model \eqref{1_trans} is considered.  Recall the existence of the budgetary restriction $\sum_{i=1}^nw_t=1$ on the relative portfolio weights.  Although it may be easy for an individual hedge fund to take a short position on an asset class, it is very hard for the aggregate of all \$2+ trillion worth of hedge funds to take a net short position on some asset class.  Therefore, we impose the restriction that asset class weightings may not take on a negative value, $w_{t,i}\ge0$ for $i\in\{1,...,n\}$.  This suggests the use of a Dirichlet distribution for the weight transitions:
\begin{equation}
w_t\sim Dir\left(\alpha\frac{w_{t-1}\circ(1+r_{PA,t-1})}{\sum_{i=1}^nw_{t-1,i}(1+r_{PA,t-1,i})}\right)\label{1_trans2}
\end{equation}
where $\alpha$ is a scalar concentration parameter controlling how much the aggregate hedge fund industry changes its investment weightings each period.  Notice how this satisfies the desired property suggested in \eqref{1_rw}:
$$
E[w_t|\mathcal{F}_{t-1}]=\frac{\boldsymbol{\alpha}}{\boldsymbol{\alpha}_0}=\frac{w_{t-1}\circ(1+r_{PA,t-1})}{\sum_{i=1}^nw_{t-1,i}(1+r_{PA,t-1,i})}
$$
where
$$
\boldsymbol{\alpha}\equiv\alpha\frac{w_{t-1}\circ(1+r_{PA,t-1})}{\sum_{i=1}^nw_{t-1,i}(1+r_{PA,t-1,i})}\;\;\text{and}\;\;\boldsymbol{\alpha}_0\equiv\sum_{j=1}^n\boldsymbol{\alpha}_j=\sum_{j=1}^n\alpha\frac{w_{t-1,j}(1+r_{PA,t-1,j})}{\sum_{i=1}^nw_{t-1,i}(1+r_{PA,t-1,i})}=\alpha
$$

Consider the observation model \eqref{1_obs}.  The purpose of the parameter $\epsilon_t$ is to pick up the variation in the hedge fund index returns $r_{HF,t}$ which is orthogonal to the palette asset returns $r_{PA,t}$, it is appropriate to consider a leptokurtic distribution due to the fat-tail property commonly exhibited by financial data first noted by Mandelbrot (1963).  Therefore, the following scale-location Student-t model is used in our analysis:
\begin{equation}
r_{HF,t}\sim t(w_t'r_{PA,t},\sigma_\epsilon^2,\nu)\label{1_obs2}
\end{equation}
By combining expressions \eqref{1_trans2} and \eqref{1_obs2}, our dynamic model is completely defined to form the foundational Dirichlet Portfolio Model (DPM):
$$
w_t\sim Dir\left(\alpha\frac{w_{t-1}\circ(1+r_{PA,t-1})}{\sum_{i=1}^nw_{t-1,i}(1+r_{PA,t-1,i})}\right)
$$
$$
r_{HF,t}\sim t(w_t'r_{PA,t},\sigma_\epsilon^2,\nu)
$$
This model can be used for estimation of latent asset weights in any portfolio where we are interested in the dynamics of weights changes due to active trading decisions.

\subsection{Filtering Method}
The use of a Dirichlet transition model, as well as the Student-t observation model has ruled out analytical solutions to this problem.  Therefore, a Sequential Monte Carlo simulation technique is used to estimate the palette asset weights.

First, a prior distribution is placed over the initial palette asset weights:
$$
w_0\sim Dir\left(\alpha_0\left(\frac{1}{n},...,\frac{1}{n}\right)'\right)
$$
where $\alpha_0$ is a scalar concentration parameter controlling initial uncertainty about the prior weight distribution.  Each asset is given equal weight in expectation with the lack of better information.  Simulating from this distribution gives a set of particles $\{w_0\}_p$, $p\in\{1,...,P\}$, characterizing the approximation.  Then, at each point in time we iteratively propagate and resample as defined in the Sequential Importance Resampling (SIR) algorithm of Rubin (1987) and Smith and Gelfand (1992).  This process produces the weight distribution estimates of interest.

First, consider the propagation step.  Let there exist a set of particles $\{w_{t-1}\}_p$ representing the distribution $p(w_{t-1}|r_{HF,t-1},r_{PA,t-1})$ from a previous iteration.  In order to find the prior distribution for the asset weights at time $t$, $p(w_t|w_{t-1},r_{PA,t-1})$, draws from the transition model for each particle $\{w_{t-1}\}_p$ are made.  This yields a set of particles $\{w_{t|t-1}\}_p$ approximating this prior distribution.  Second, consider the resampling step.  Given the set of particles $\{w_{t|t-1}\}_p$ approximating the prior distribution, they are resampled with respect to their relative likelihoods given by $\omega_t=p(r_{HF,t}|w_t,r_{PA,t})$.  This set of resampled particles $\{w_t\}_p\equiv\{w_{t|t}\}_p$ will therefore approximate the desired distribution $p(w_t|r_{HF,t},r_{PA,t})$.

Using these results, forms for all of the probability distributions in the SIR algorithm are fully specified.  First, the `Step 1' propagation step is defined:
$$
p(w_t|w_{t-1}^{(p)},r_{PA,t-1})\equiv Dir\left(\alpha\frac{w_{t-1}^{(p)}\circ(1+r_{PA,t-1})}{\sum_{i=1}^nw_{t-1,i}^{(p)}(1+r_{PA,t-1,i})}\right)
$$

Second, the `Step 2' normalized importance weights are computed from the observation model $r_{HF,t}\sim t(w_t'r_{PA,t},\sigma_\epsilon^2,\nu)$ where
$$
p(r_{HF,t}|w_t,r_{PA,t})=\frac{\Gamma(\frac{\nu+1}{2})}{\Gamma(\frac{\nu}{2})\sqrt{\pi\nu\sigma_\epsilon^2}}\left(1+\frac{1}{\nu}\frac{(r_{HF,t}-w_t'r_{PA,t})^2}{\sigma_\epsilon^2}\right)^{-\frac{\nu+1}{2}}
$$
Therefore, the importance weights are given by:
$$
\omega_t^{(p)}\equiv\frac{p(r_{HF,t}|w_t^{(p)},r_{PA,t})}{\sum_{\phi=1}^Pp(r_{HF,t}|w_t^{(\phi)},r_{PA,t})}=\frac{\left(\nu\sigma_\epsilon^2+\left(r_{HF,t}-w_t^{(p)\prime}r_{PA,t}\right)^2\right)^{-\frac{\nu+1}{2}}}{\sum_{\phi=1}^P\left(\nu\sigma_\epsilon^2+\left(r_{HF,t}-w_t^{(\phi)\prime}r_{PA,t}\right)^2\right)^{-\frac{\nu+1}{2}}}
$$
Finally, the initial palette asset weight distribution is set:
$$
p(w_0)\equiv Dir\left(\alpha_0\left(\frac{1}{n},...,\frac{1}{n}\right)'\right)
$$
Note that if there exists better information about the distribution $p(w_0)$, the use of that will naturally lead to superior and more appropriate results.

Now, we substitute in these developed forms for the Dirichlet Portfolio Model to get the fully specified DPM Estimation Algorithm in Figure \ref{1_algDPM}.
\begin{figure}
\noindent\fbox{
\begin{minipage}[h]{0.968\textwidth}
\begin{center}
\textbf{DPM Estimation Algorithm}
\end{center}

Initialize: Sample prior weights from
$$
w_0^{(p)}\sim Dir\left(\alpha_0\left(\frac{1}{n},...,\frac{1}{n}\right)'\right)
$$
Iterate:

\hspace{15pt} Step 1: Propagate new asset weights from
$$
w_t^{(p)}\sim Dir\left(\alpha\frac{w_{t-1}^{(p)}\circ(1+r_{PA,t-1})}{\sum_{i=1}^nw_{t-1,i}^{(p)}(1+r_{PA,t-1,i})}\right)\text{ for }p=1,...,P
$$
\hspace{15pt} Step 2: Resample asset weights from
$$
w_t^{(p)}\sim Mult_P\left(\left\{\omega_{t}^{(\phi)},w_t^{(\phi)}\right\}_{\phi=1}^P\right)
$$
\hspace{15pt} where the normalized importance weights are given by
$$
\omega_t^{(p)}=\frac{\left(\nu\sigma_\epsilon^2+\left(r_{HF,t}-w_t^{(p)\prime}r_{PA,t}\right)^2\right)^{-\frac{\nu+1}{2}}}{\sum_{\phi=1}^P\left(\nu\sigma_\epsilon^2+\left(r_{HF,t}-w_t^{(\phi)\prime}r_{PA,t}\right)^2\right)^{-\frac{\nu+1}{2}}}
$$

\end{minipage}
}
\caption{\textsl{\small{DPM Estimation Algorithm}}}
\label{1_algDPM}
\end{figure}

\subsection{Conditionally Normal Approximation}

Due to the non-Gaussian nature of this multivariate compositional model, the posterior distributions of the asset weights cannot be solved for in analytical closed form.  Therefore, sequential Monte Carlo methods are employed to numerically approximate these distributions.  Another approach to solving this problem is to approximate the Dirichlet errors by a multivariate Gaussian distribution replicating the first two moments at each step time.  So, just as was done in the above solution, the error distributions of the transitions must be reparametrized at each time step based upon the estimation results from the previous step.

First, consider the transition model from the DPM in \eqref{1_trans2}.  It can be shown that the first two moments of $w_t$ are:
$$
\mu_t\equiv E[w_t|\mathcal{F}_{t-1}]=\frac{w_{t-1}\circ(1+r_{PA,t-1})}{\sum_{i=1}^nw_{t-1,i}(1+r_{PA,t-1,i})}
$$
and
$$
Cov[w_{t,i},w_{t,j}|\mathcal{F}_{t-1}]=\frac{\boldsymbol{\alpha}_i(\boldsymbol{\alpha}_0I_{i=j}-\boldsymbol{\alpha}_j)}{\boldsymbol{\alpha}_0^2(\boldsymbol{\alpha}_0+1)}
$$
where
$$
\boldsymbol{\alpha}_i\equiv \alpha\frac{w_{t-1,i}(1+r_{PA,t-1,i})}{\sum_{i=1}^nw_{t-1,i}(1+r_{PA,t-1,i})}\quad\text{and}\quad\boldsymbol{\alpha}_0=\sum_i\boldsymbol{\alpha}_i=\alpha
$$
from above.  So, it can be shown that:
$$
\Sigma_{t,i,j}\equiv Cov[w_{t,i},w_{t,j}|\mathcal{F}_{t-1}]=\frac{w_{t-1,i}(1+r_{PA,t-1,i})\left(\xi_{t-1} I_{i=j}-w_{t-1,j}(1+r_{PA,t-1,j})\right)}{\xi_{t-1}(\alpha+1)}
$$
and
$$
\xi_{t-1}=\sum_{k=1}^nw_{t-1,k}(1+r_{PA,t-1,k})
$$

Then, using this, the original DPM Model can be approximately rewritten into the Conditionally Normal Dirichlet Portfolio Model (CN-DPM) assuming a Gaussian observational distribution:
$$
w_t\sim N\left(\frac{w_{t-1}\circ(1+r_{PA,t-1})}{\sum_{i=1}^nw_{t-1,i}(1+r_{PA,t-1,i})},\left[\frac{w_{t-1,i}(1+r_{PA,t-1,i})\left(\xi_{t-1} I_{i=j}-w_{t-1,j}(1+r_{PA,t-1,j})\right)}{\xi_{t-1}(\alpha+1)}\right]_{i,j}\right)
$$
$$
r_{HF,t}\sim N(w_t'r_{PA,t},\sigma_\epsilon^2)
$$
Note that although the observational distribution may not be best modeled by a Gaussian form, it is a necessary simplification to use the results from Kalman (1960) to solve the dynamic model analytically.  Using the above form, the latent weights are solved for in the Conditionally Normal Dirichlet Portfolio Model in Figure \ref{1_algCNDPM}.
\begin{figure}
\noindent\fbox{
\begin{minipage}[t]{0.968\textwidth}
\begin{center}
\textbf{CN-DPM Estimation Algorithm}
\end{center}

Initialize: Set initial weight distribution
$$
p(w_0)=N\left(\mu_0,\Sigma_0\right)
$$
Iterate:

\hspace{15pt} Step 1: Compute the prior weight distribution using the transition model
$$
p(w_t|\mathcal{F}_{t-1})=N\left(\mu_{t|t-1},\Sigma_{t|t-1}\right)
$$
$$
\mu_{t|t-1}=\frac{w_{t-1}\circ(1+r_{PA,t-1})}{\sum_{i=1}^nw_{t-1,i}(1+r_{PA,t-1,i})}
$$
$$
\Sigma_{t|t-1}=\Sigma_{t-1|t-1}+\left[\frac{w_{t-1,i}(1+r_{PA,t-1,i})\left(\xi_{t-1} I_{i=j}-w_{t-1,j}(1+r_{PA,t-1,j})\right)}{\xi_{t-1}(\alpha+1)}\right]_{i,j}
$$
$$
\xi_{t-1}=\sum_{k=1}^nw_{t-1,k}(1+r_{PA,t-1,k})
$$
\hspace{15pt} Step 2: Compute the posterior weight distribution using the observation model
$$
p(w_t|\mathcal{F}_t)=N\left(\mu_{t|t},\Sigma_{t|t}\right)
$$
$$
\mu_{t|t}=\mu_{t|t-1}+K_t(r_{HF,t}-\mu_{t|t-1}'r_{PA,t})
$$
$$
\Sigma_{t|t}=\left(I-K_tr_{PA,t}'\right)\Sigma_{t|t-1}
$$
\hspace{15pt} where the optimal Kalman Gain value is given by
$$
K_t=\left(\Sigma_{t|t-1}r_{PA,t}\right)\left(r_{PA,t}'\Sigma_{t|t-1}r_{PA,t}+\sigma_\epsilon^2\right)^{-1}
$$

\end{minipage}
}
\caption{\textsl{\small{CN-DPM Estimation Algorithm}}}
\label{1_algCNDPM}
\end{figure}

\section{Alternative Approaches}

We briefly review some alternative estimation techniques that either have been used, or could be used similarly to the Dirichlet Portfolio Model.

\subsection{Rolling Window OLS}
The general setup of these regression models is as follows:
$$
r_{HF,\tau}=w_t'r_{PA,\tau}+\epsilon_\tau,\quad\epsilon\sim N(0,\sigma^2)
$$
where $\tau\in\mathrm{T}\equiv\{t-k,t-k+1,...,t\}$ for a $k$-sized window.  The classic OLS estimator is given by $\hat{w}_t=\left(r_{PA,\mathrm{T}}'r_{PA,\mathrm{T}}\right)^{-1}r_{PA,\mathrm{T}}'r_{HF,\mathrm{T}}$.  While this is a simplified first approach, it suffers from some major problems.  First, this is a static model for the estimated asset weights, and therefore makes the incorrect assumption that the weights are constant over the estimation time period.  Second, a window size $k$ must be chosen, therefore having to deal with the trade-off of using more data to obtain better estimates but decreasing the relative importance of more recent observations.  Lastly, the most apparent problem is the lack of the $\sum_{i=1}^nw_{t,i}=1$ restriction.  Nevertheless, a more appropriate estimator using this portfolio normalization constraint can be constructed.  We have the following setup:

$$
\bar{w}_t=\argmin_{w_t}\sum_{\tau\in\mathrm{T}}\left(r_{HF,\tau}-w_{t}'r_{PA,\tau}\right)^2\quad\text{where}\quad\sum_{i=1}^nw_{t,i}=1
$$
This can be solved by Constrained Least Squares (CLS) from Chipman and Rao (1964) and Tintner (1952).  In the context of this problem, estimates are obtained by:
$$
\bar{w}_t=\hat{w}_t-\left(r_{PA,\mathrm{T}}'r_{PA,\mathrm{T}}\right)^{-1}\mathbf{1}\left(\mathbf{1}'\left(r_{PA,\mathrm{T}}'r_{PA,\mathrm{T}}\right)^{-1}\mathbf{1}\right)^{-1}\left(\mathbf{1}'\hat{w}_t-1\right)
$$

Note that although the above solution does place a normalizing restriction on the sum of the estimated weights, it still allows for individual weights to take any real value.  That is, an estimated weight of 120 or -80 could be obtained, thereby implying an unrealistic 12,000\% or -8,000\% weight on that asset class.  This explosive scaling effect happens widely in the presence of multicollinearity in the explanatory variables.  Due to the prevalence of this in asset returns, the undesired scaling issue can be avoided by imposing the ``no short selling'' assumption of the DPM on the above CLS.  That is, constrain the CLS with non-negativity: $w_{t,i}\ge 0,\quad\forall i\in\{1,\dots,n\}$.  Let us refer to this setup as the Inequality Constrained Least Squares (ICLS) method similar to Judge and Takayana (1966) and Liew (1976).  This is easily solved via quadratic optimization.

Herein, the original OLS rolling regression approach will not be considered due to its gross misspecification for this problem.  Instead, the CLS and ICLS approaches will be explored due to their increased suitability.

\subsection{Na\"{\i}ve Kalman Filtering}
Amenc et al outline an approach for using Bayesian inference to solve the dynamic state space model.  The model is set up as follows:
$$
r_{HF,t}=w_t'r_{PA,t}+\epsilon_t,\quad\epsilon_t\sim N(0,\sigma^2_\epsilon)
$$
$$
w_t=w_{t-1}+\eta_t,\quad\eta_t\sim N(0,Q)
$$

This is a classic state space model which is analytically solvable via the Kalman Filter.  Although this approach correctly identifies the problem as a dynamic model, it lacks the portfolio normalization constraint $\sum_{i=1}^nw_{t,i}=1$.  So, similar to the least-squares based techniques detailed above, we propose more suitable methods by adding in this constraint.
\begin{enumerate}
\item \textbf{Constrained Kalman Filtering via Restricted Covariance Structure}\\
Consider the constraint $\sum_{i=1}^nw_{t,i}=1$.  It can be shown that
$$
\sum_{i=1}^nCov\left(w_{t,j},w_{t,i}\right)=0,\quad\forall j\in\{1,\dots,n\}
$$
Therefore, we can obtain a constrained estimation structure by choosing the initial weight uncertainty matrix $\Sigma_0$ and transition innovation matrix $Q$ such that:
\begin{equation}
\Sigma_0\mathbf{1}=\mathbf{0},\, Q\mathbf{1}=\mathbf{0}\quad\text{and by symmetry}\quad \Sigma_0'\mathbf{1}=\mathbf{0},\, Q'\mathbf{1}=\mathbf{0}\label{1_covRes}
\end{equation}

Now, with the lack of better information, assign to $Var(w_{t,i})$ the average unconditional variance implied by the $\alpha$ parameter from the DPM for consistency purposes.  Then, let
$$
Cov(w_{t,i},w_{t,j})=-\frac{Var(w_{t,i})}{n-1},\quad\forall j\neq i,
$$
thereby satisfying the above restriction and creating a spherical covariance structure.  Nevertheless, if there does exist information about a more suitable covariance structure, but it does not satisfy the above properties, we can project the original covariance matrices onto the space satisfying those constraints:
$$
\Sigma_0^P=\left(I-\Sigma_0\mathbf{1}\left(\mathbf{1}'\Sigma_0\mathbf{1}\right)^{-1}\mathbf{1}'\right)\Sigma_0\quad\text{and}\quad Q^P=\left(I-Q\mathbf{1}\left(\mathbf{1}'Q\mathbf{1}\right)^{-1}\mathbf{1}'\right)Q
$$

\item \textbf{Constrained Kalman Filtering via State Projection}\\
Chia (1985) and Simon and Chia (2002) detail a method to first derive the unconstrained state estimate and then project it onto the constraint surface.  This can easily be applied in the context of this application.  When computing the posterior distribution of the weights, we can arrive at the projected distribution $N(\mu^P_{t|t},\Sigma^P_{t|t})$ by first computing the unconstrained solution $N(\mu_{t|t},\Sigma_{t|t})$ in the classic manner, and then projecting via:
$$
\Upsilon_t=\Sigma_{t|t}\mathbf{1}\left(\mathbf{1}'\Sigma_{t|t}\mathbf{1}\right)^{-1}\qquad\mu^P_{t|t}=\mu_{t|t}-\Upsilon_t\left(\mathbf{1}'\mu_{t|t}-1\right)\qquad\Sigma^P_{t|t}=\left(I-\Upsilon_t\mathbf{1}'\right)\Sigma_{t|t}
$$
Conveniently, for calculating the prior distribution/forecasts for the weights, our transition function is already normalized with respect to the posterior weights from the previous period, thereby already projecting into the constrained space.  Again, we assign to $Var(w_{t,i})$ the average unconditional variance implied by the $\alpha$ parameter from the DPM for consistency purposes.

Note that the estimation error covariance $\Sigma_0$ of this method is always going to be greater than or equal to that obtained by using the restricted covariance matrix method (Ko and Bitmead, 2007).  This is because in the restricted covariance matrix method, the transition innovation covariance matrix $Q$ is assumed to be the true process noise covariance, thereby resulting in the optimal state estimates for the system.  However, in the projection method, the transition innovation covariance matrix $Q$ may be inconsistent with the estimated transition innovation process.  Nevertheless, if $Q$ satisfies \eqref{1_covRes}, then both these methods are equivalent.
\end{enumerate}

Also, it is possible to incorporate inequality constraints into the Kalman Filtering approach as we did for the ICLS.  Gupta and Hauser (2007) detail a method to do so using quadratic optimization.  We do not implement this approach here since generally the Constrained Kalman solutions above produce estimation which is consistent with the desired non-negativity constraints, thereby negating the need to implement them in the estimation procedure.

Note that the CN-DPM presented above is analytically solvable via a modified Kalman filtering approach, however it incorporates an approximation for the Dirichlet compositional structure suggested by the DPM.  This importantly requires redefined error distributions at each period.  As well, note that the CN-DPM is a special case of the Constrained Kalman Filtering via Covariance Structure class of models due to it's compliance with the covariance restrictions.  Furthermore, due to the period-by-period redefined error distributions, it also allows for the non-negativity constraint.

Again, due to the na\"{\i}ve Kalman Filter's misspecification for this problem, we only explore the Constrained Kalman via Covariance Structure (CKalCov) and State Projection (CKalProj) methods.

\section{Simulated Portfolio Trading Comparison}

We now compare the DPM and CN-DPM with the presented alternative estimation techniques on various simulated portfolio environments to motivate the effectiveness of the procedure.

\subsection{Simulated Assets \& Trading}

First, sets of simulated monthly asset returns are developed under the following example model:
$$
r_{i,t}\sim t(\mu_i, \sigma_i^2, \nu_i)
$$
where
$$
\mu_i\sim N(0.007,0.003^2),\quad\sigma_i^2\sim IG(2.5,0.004),
$$
$$
\nu_i\sim IU(0,0.5),\quad\Sigma_i\sim IW(\mathbf{I},8)
$$
with contemporary correlation induced by a Gaussian copula having correlation implied by $\Sigma_i$.  This example parametrization was motivated by Gelman and Hill (2006).  Nevertheless, we will later demonstrate that the results also hold with real asset returns.  For the following simulations, the case of 6 investable assets is considered.
  
Using these simulated assets, there are various ways to construct time series of portfolio weights.  Let us first construct simulated portfolio asset weights using the previously motivated random-walk process from \eqref{1_trans2}.  Using these simulated weights, the simulated time series of hedge fund returns is constructed via $r_{HF,t}\sim t(w_t'r_t,\sigma_\epsilon^2,\nu_{HF})$.  A Student-t distribution is used here since it is reasonable to observe that returns which are orthogonal to the set of included explanatory assets can potentially be very ``fat-tailed'' in nature due to some trading activities such as market-making, high frequency trading, highly illiquid asset pricing, etc.

The objective is to track the weights on each of these assets, but due to the unobservable nature of the true portfolio weights, the accuracy of the weight predictions can never be observed in the real world setting.  Therefore, to gain a proxy of how close these weights are being estimated, we can create the forecasted set of weights from the model and then determine how close we are to the one-step-ahead returns from the hedge fund index.  That is, we want to minimize the following error:
$$
\epsilon_t = r_{HF,t} - E[w_t|\mathcal{F}_{t-1}]'r_t
$$
There are various measures of accuracy for this estimation.  We explicitly define the following four measures for use throughout the remainder of the paper:

\begin{table}[H]
\centering
\small
\begin{tabular}{cc}
\textbf{Measure Name}&\textbf{Expression}\\
\hline
Forecasted Root Mean Squared Error (F-RMSE)&$\sqrt{\frac{1}{T}\sum_{t=1}^T(r_{HF,t} - E[w_t|\mathcal{F}_{t-1}]'r_t)^2}$\\
Forecasted Mean Absolute Error (F-MAE)&$\frac{1}{T}\sum_{t=1}^T\left|r_{HF,t} - E[w_t|\mathcal{F}_{t-1}]'r_t\right|$\\
Forecasted Pearson Correlation (F-Corr)&$corr\left(r_{HF,t},E\left[w_t|\mathcal{F}_{t-1}\right]'r_t\right)$\\
Forecasted $R^2$ (F-$R^2$)&$1-\frac{\sum_{t=1}^T\left(r_{HF,t} - E[w_t|\mathcal{F}_{t-1}]'r_t\right)^2}{\sum_{t=1}^T\left(r_{HF,t} - \bar{r}_{HF}\right)^2}$\\
\end{tabular}
\end{table}

Note that this F-RMSE value is exactly the same as the ``tracking error'' concept used commonly in portfolio management to describe how close the returns of a portfolio track to the returns of a given benchmark index.  In our case, the benchmark index is simply the hedge fund index of interest.

\begin{figure}[ht]
\centering
\includegraphics[width=0.55\textwidth]{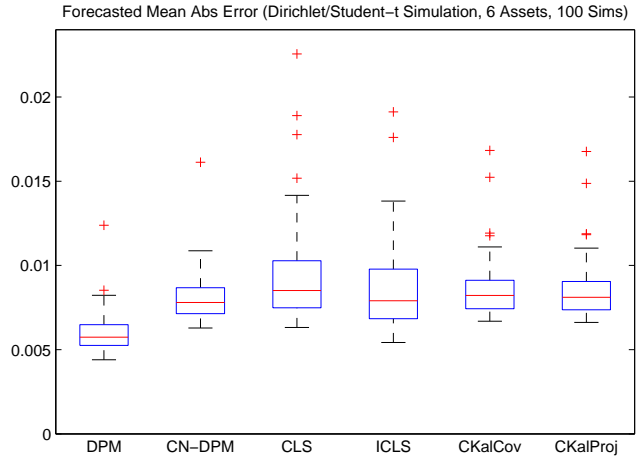}
\caption{\textsl{\small{Simulation Results -- Forecasted Return MAE}}}
\label{1_foreMAE}
\end{figure}

We run 100 simulations and estimate the portfolio weights using the DPM, CN-DPM, CLS, ICLS, CKalCov, and CKalProj.  First, we compare the simulation results using Forecasted Mean Absolute Error in Figure \ref{1_foreMAE}.  The DPM and CN-DPM procedures produce smaller and more precise forecasted mean absolute deviation values, and therefore more accurate forecasted returns than the other methods.  The constrained Kalman Filter methods, CKalCov and CKalProj, generally produce the next best best results, however, with a median forecasted MAE of $0.0082$ and $0.0081$ versus the DPM's value of $0.0057$, the Kalman Filters perform about $43\%$ worse.  This is compared to using rolling window CLS and ICLS, which give median forecasted MAE values of $0.0085$ and $0.0079$, $49\%$ and $39\%$ worse than the DPM.  As well, the CN-DPM has a median forecasted MAE of $0.0078$, second to the DPM.

\begin{figure}[ht]
\centering
\includegraphics[width=0.55\textwidth]{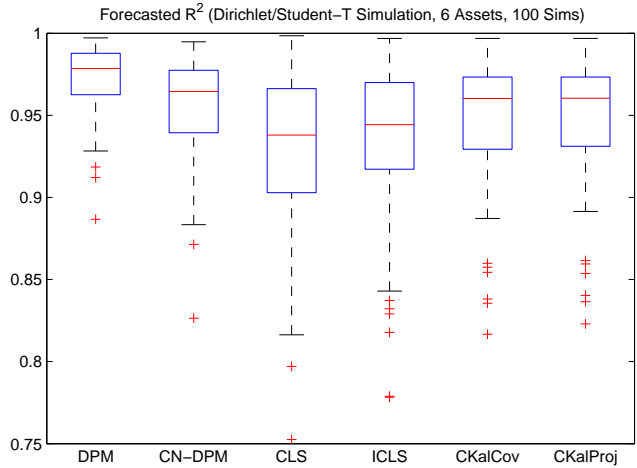}
\caption{\textsl{\small{Simulation Results -- Forecasted Return $R^2$}}}
\label{1_foreR2}
\end{figure}

Another way to compare these estimation procedures is the Forecasted Coefficient of Determination, F-$R^2$.  Naturally, this does not consider scale, as the MAE measure does, but it is useful to consider the proportion of variation in the hedge fund returns explained by the forecasted model.  Examining the plot in Figure \ref{1_foreR2}, the DPM produces much stronger forecasted $R^2$ values than the other methods, thus supporting its more accurate asset weight estimation.  The DPM and CN-DPM produce median forecasted $R^2$ values of $0.979$ and $0.965$, respectively, as compared to rolling CLS and ICLS rolling with $0.938$ and $0.944$, and the CKalCov and CKalProj with $0.960$ and $0.960$.  Similar plots can be created for the F-RMSE and F-Corr measures.  These plots display similar results as those shown here.

\subsection{Increasing Number of Assets}

Since the dimension of explanatory assets has the potential to grow large, we explore the effect of increasing the number of investable assets.  Below, the same simulations are ran, but while increasing the number of investable assets in the simulated hedge fund index construction.

\begin{figure}
\centering
\includegraphics[width=0.55\textwidth]{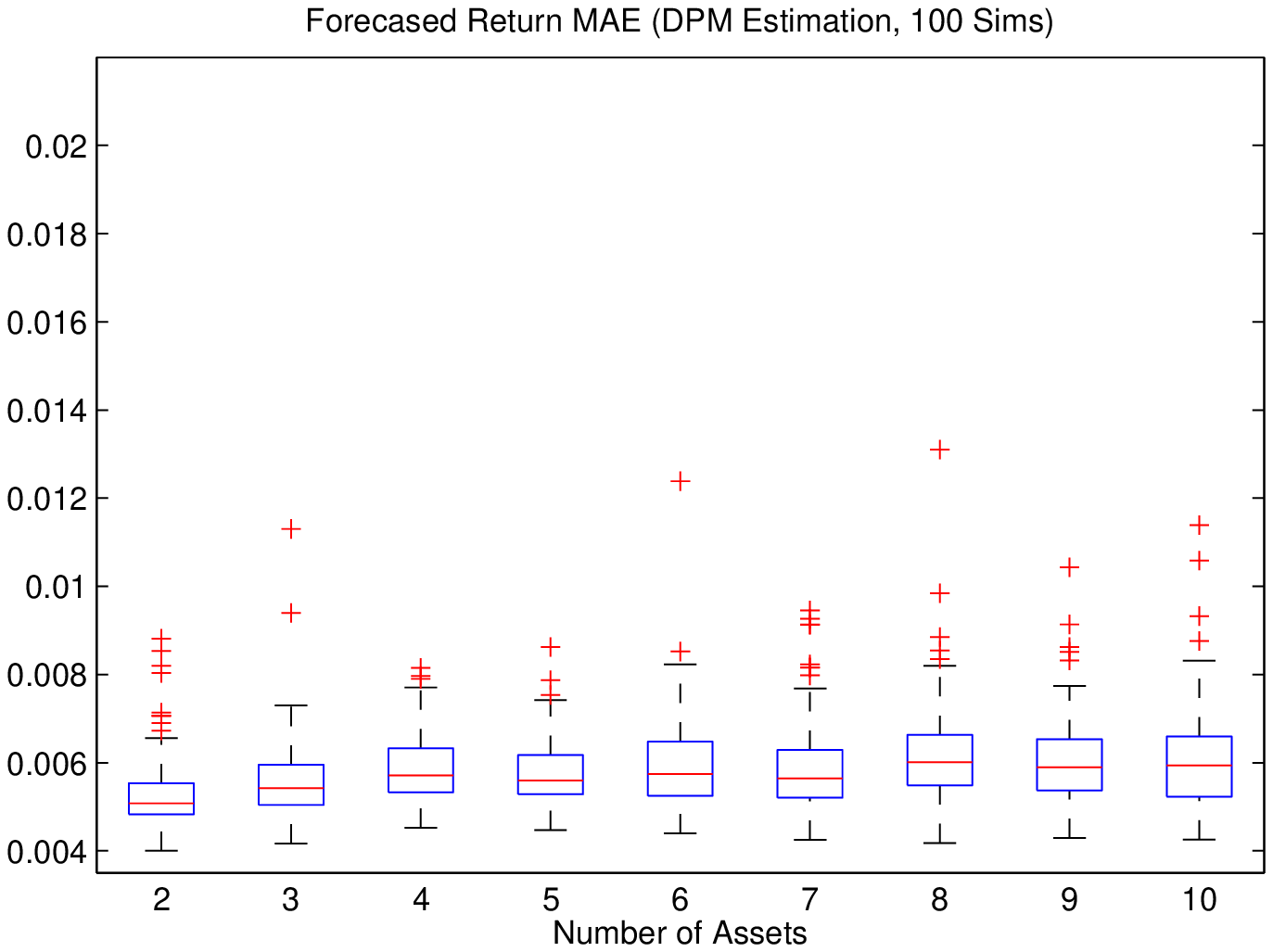}
\caption{\textsl{\small{DPM Forecasted Return MAE vs. Number of Assets}}}
\label{1_MAE_DPM}
\vspace{\baselineskip}
\includegraphics[width=0.55\textwidth]{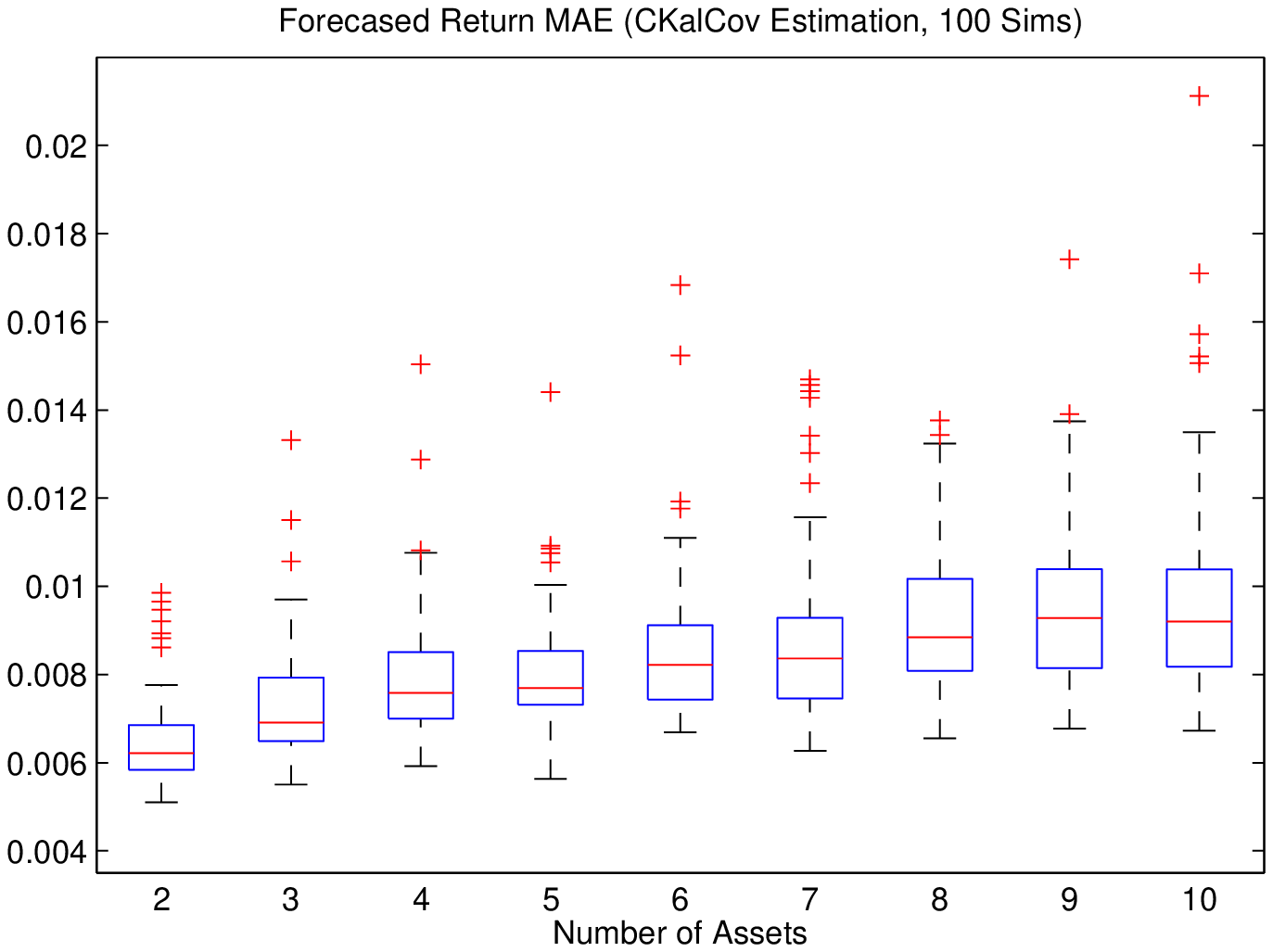}
\caption{\textsl{\small{CKalCov Forecasted Return MAE vs. Number of Assets}}}
\label{1_MAE_CKalCov}
\end{figure}

In Figure \ref{1_MAE_DPM}, the DPM procedure is used to estimate weights on the simulated assets, and then those weights are used to construct the forecasted return MAE values for each simulation.  As expected, as the number of assets increases, the forecasted return MAE increases.  However, considering that the dimension of the estimation space is increasing, and therefore the potential estimation outcomes are exponentiating, the forecasted return MAE values do not degrade unreasonably.  As well, although adding additional assets increases forecasted return MAE, as expected, we do observe that each new asset has a decreasing marginal effect on this accuracy measure.

For comparison, let us consider one of the constrained Kalman Filtering procedures from above, the CKalCov.  Figure \ref{1_MAE_CKalCov} uses this method to estimate weights on the simulated assets, and then constructs the forecasted return MAE values for each simulation.  Here, the forecasted return MAE values degrade at a much faster rate than with the DPM.  In fact, this relationship is unfortunately much more linear as the number of assets increase.  Furthermore, at large numbers of assets, there are many more extreme values for inaccurate forecasted return MAE.

Although they are not exhibited here, similar plots for CLS and ICLS demonstrate even less accurate results.  Therefore, the DPM is increasingly favored when considering a sizable potential space of investable assets.  In the real world, we observe that hedge funds can invest in a vast number of potential assets, thus the DPM becomes an even more useful tool for estimating weights on that asset set.

\subsection{Real Assets \& Model Hedge Fund Example}

To further motivate the suitability of the DPM, we construct a realistic hedge fund trading model similar to the one proposed in Khandani and Lo (2007).  Then, the resulting asset weights are used to create a model hedge fund return series.  We then use this return series, along with the portfolio asset returns, in the estimation procedures to obtain estimates for the asset weights.  Finally, these estimated weights are compared to the true asset weights to infer accuracy.  As the portfolio assets, we use the daily returns from the four largest sectors in S\&P 500 Index (Technology, Financials, Health Care, and Energy) from January 4, 2010 to February 13, 2013.

\begin{figure}
\centering
\includegraphics[width=0.95\textwidth]{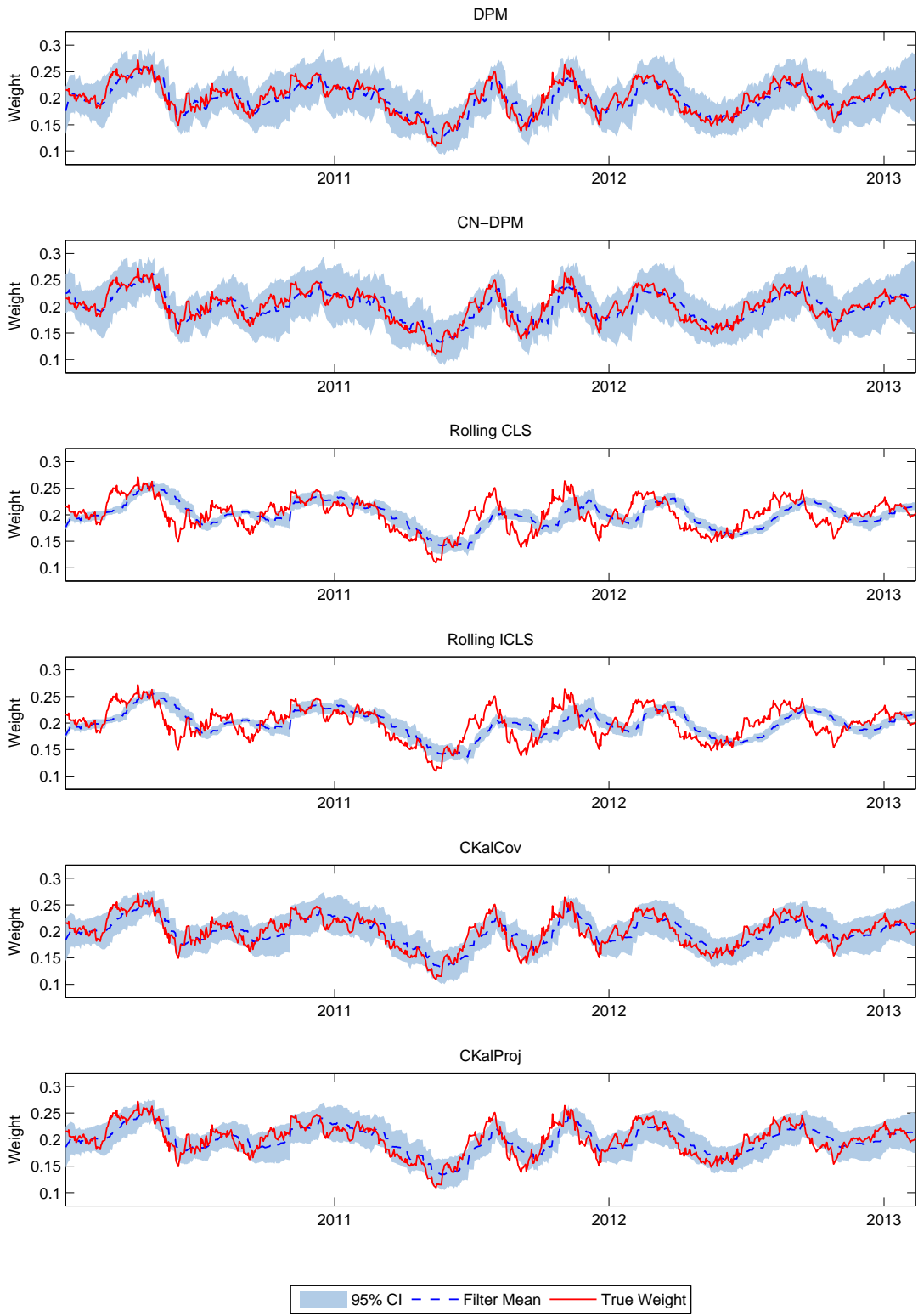}
\caption{\textsl{\small{Estimated Weight Accuracy Comparison}}}
\label{1_realWeightCI}
\end{figure}

Let us construct the model trading strategy as follows.  Given a set of $N$ equity sectors, consider a strategy where these are held proportional to their market capitalization $\tilde{w}$, but these weights are decreased or increased proportional to previous over or under-performance relative to their average.  That is, the sectors which have previously over-performed are relatively under-weighted, while the sectors which have previously under-performed are relatively over-weighted.  This is a form of ``contrarian'' strategy by under-weighting yesterday's winners and over-weighting yesterday's losers.  For our example, we use the aggregated sum of daily returns over the last 30 days for each asset ($R_{t-1,i}\equiv\sum_{\tau=t-30}^{t-1}r_{\tau,i}$) when constructing our over/under-weight values.  Specifically, the following asset weights are constructed:
$$
w_{t,i}=\tilde{w}_{t,i}-\left(R_{t-1,i}-R_{t-1,m}\right),\quad R_{t,m}\equiv\frac{1}{N}\sum_{i=1}^NR_{t-1,i}
$$
These are taken as our ``true'' weights used to construct the hedge fund return series.  Now, to compare the accuracy of our procedures, we look at the estimation of our hedge fund's weight process.  Figure \ref{1_realWeightCI} exhibits the resulting weight point-estimates and confidence intervals (or respective Bayesian credible intervals) for the S\&P 500 Health Care Sector.

Visually, the DPM does the best job of tracking the true weight time series with an MAE of 1.16\%, while the CN-DPM comes in a close second with 1.23\%.  For comparison, the OLS methods perform 43-75\% worse, while the Kalman methods perform 12-21\% worse, across the different component assets.  As well, not only does the DPM accurately estimate the underlying weight process, but also it adjusts to large changes in the weights very quickly.

The rolling CLS and ICLS, as expected, take quite a few periods to adjust to large changes in the weights since the importance of each observation is given equal weight in determining the resulting weight estimates.  Therefore, a new portfolio return observation does not have a large impact on the weight estimation, especially if a large window size is used.  However, if too small a window size is used, poor estimates and confidence intervals are obtained since the estimation sample is too small.  Furthermore, since there is no intertemporal structure placed on weight transitions, we can obtain unreasonably large jumps in the estimated weights due to multicollinearity in the palette asset returns.

The constrained Kalman filtering methods do a better job, however they still do not track the true weight series as accurately as the DPM.  This is especially evident when looking at quick changes in the weight values, however it is not as severe as compared to the rolling CLS and ICLS since the structure of the Kalman Filter allows for more appropriate updating of the weight estimation after obtaining a new portfolio return observation.  Furthermore, similar to the results of the previous subsection, as the number of component assets increase, the DPM performs increasingly better, relative to the other methods.

\section{Empirical Results \& Comparison}

\subsection{Hedge Fund Data}

We apply the DPM estimation methodology to monthly return data for the Hedge Fund Research Fund Weighted Composite Index (HFRIFWC) from January 1995 to October 2012.  These return values are reported net of individual fund managers' fees.  Since this index is constructed by compiling self reported hedge fund returns from individual managers, there are a few potential biases to identify in the data.  First, there is no requirement that fund managers report their monthly returns, therefore only a subset of all funds report into these aggregated indices.  Some hedge fund strategies have a maximum asset size which can be effectively invested, thereby creating a cap on the total fund size.  In this case, some funds who have reached their maximum size may have no incentive to report their returns.  This creates a downward bias in the self reported return indices if these missing funds are outperforming the average.  On the other hand, there is a selection survivorship bias in the self reported returns since only the funds that are continuing to operate and therefore have not experienced large losses are reporting returns.  This creates an upward bias in these self reported returns.  Nevertheless, a large portion of fund managers do report returns, mainly for advertising purposes.  Therefore, this aggregated index is the best proxy available for the whole hedge fund industry's returns to investors.  So, we can reasonably use this return series in estimating the weights on our set of palette assets in the following sections.

\subsection{Palette Assets \& Parametrization}

Since the universe of investable assets is quite large, it is very difficult to estimate weights on the complete set.  However, since the goal is to estimate the invested weights on the assets that the value-weighted aggregate of hedge funds is invested in, this problem is simplified significantly.  Instead of trying to estimate weights on each and every single investable asset, we can estimate the weights on portfolios of assets (or indices) representing broad classes of assets (e.g. US equity, emerging market equity, high yield bonds, etc.).  Since the total hedge fund industry is so large, it is reasonable to make the assumption that the value weighted aggregate of hedge funds is invested in each of these broad asset classes in an approximate weighting scheme that is similar to the asset class's value weights.  Therefore, instead of trying to estimate the weights on a potentially infinite set of individual assets, it is possible to estimate these exposures on a small subset of asset classes.  Due to the smaller size of the number of asset classes, the dimension of this problem is significantly reduced, and therefore the resulting estimates are dramatically improved.

Herein, the following indices, similar to those used in Fung and Hsieh (1997), are used as a proxy for the asset classes that the hedge fund industry is investing in. Table \ref{1_palette} enumerates the asset class list and color key that will be used throughout the remainder of this paper.  Note that it is not necessary to restrict ourselves to this specific set of asset classes.  The methodology described above can be applied to any asset set of interest.

\begin{table}
\centering
\small
\begin{tabular}{|c|l|}
\hline
\textbf{Color}&\multicolumn{1}{c|}{\textbf{Asset Class Name}}\\
\hline
\cellcolor[rgb]{0.5,0,0}&Barclays Municipal Bond Index\\
\cellcolor[rgb]{1,0,0}&Barclays Short Term Treasury Bond Index\\
\cellcolor[rgb]{1,0.6667,0}&Barclays Corporate High Yield Bond Index\\
\cellcolor[rgb]{0.6667,1,0.3333}&Deutsche Bank US Dollar Long Futures Index\\
\cellcolor[rgb]{0.3333,1,0.6667}&Dow Jones - UBS Commodity Index\\
\cellcolor[rgb]{0,0.6667,1}&MSCI Emerging Markets Index\\
\cellcolor[rgb]{0,0,1}&MSCI EAFE Index (Europe, Australasia, Far East)\\
\cellcolor[rgb]{0,0,0.5}&MSCI US Equity Index\\
\hline
\end{tabular}
\normalsize
\caption{\textsl{\small{Asset Class ``Palette''}}}
\label{1_palette}
\end{table}

As well, the parametrization of the error distributions must be specified.  For the transitional portion of the model, the Dirichlet errors are parametrized by the multivariate concentration parameter $\boldsymbol{\alpha}$.  Intuitively, since the Dirichlet distribution is the conjugate prior of the multinomial distribution, this $\boldsymbol{\alpha}$ vector can be viewed as pseudo-counts for the prior distribution on the transitioned state of asset weights.  In other words, it is the relative weight of the prior when updating with the return observation to obtain the posterior asset weight distribution at a given time period.  Recall that $\sum_i\boldsymbol{\alpha}_i=\alpha$, therefore the relative weight on the prior is given solely by $\alpha$, where the weight on a single observation is 1.  This conveniently allows us to effectively quantify the influence that a single new observation has on each step in the estimation procedure as $1/(1+\alpha)$.

\begin{figure}[ht]
\centering
\includegraphics[width=0.65\textwidth]{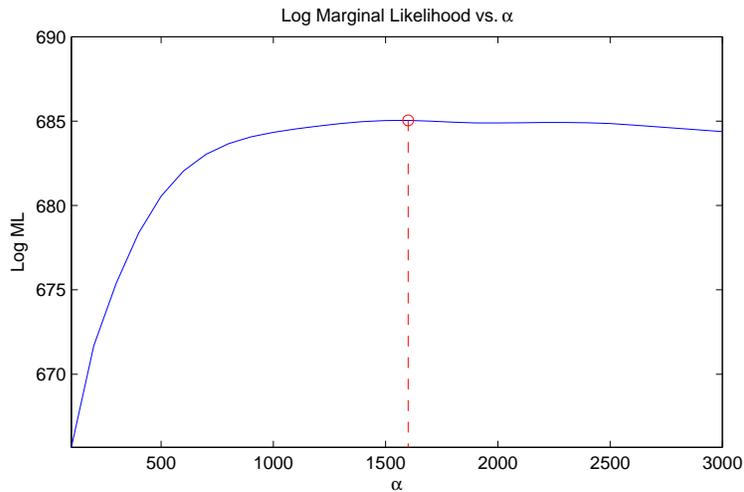}
\caption{\textsl{\small{Choice of Parameter $\alpha$}}}
\label{1_alphaChoice}
\end{figure}

Nevertheless, it is convenient to perform Bayesian model comparison via Bayes Factors in this application.  As essentially a likelihood ratio between competing model parametrizations, we simply can compute the marginal likelihoods for each $\alpha$ and choose the largest value.  Figure \ref{1_alphaChoice} shows the log marginal likelihood values for the DPM procedure estimated at various values for $\alpha$.  The maximum value is achieved at $\alpha=1600$.  We note that, deviations around $\pm 70\%$ of this choice of $\alpha$ do not change the following results materially.

Lastly, the observational portion of the model needs to be specified.  Therefore, the variance $\sigma_\epsilon^2$ and the degrees of freedom $\nu$ of the Student-t distribution need to be chosen.  Here, there is more flexibility in the choice of these parameters based upon the selection of palette assets and beliefs about the unexplained portion of returns.  The better the palette assets represent the investable universe, the smaller the choice of $\sigma_\epsilon^2$.  As well, the more likely it is to observe extreme values in the unexplained returns, the smaller the value of $\nu$ is desired.  Here, $\sigma_\epsilon^2=0.01$ and $\nu=6$ are used since the above palette assets represent the investable universe well, but can allow for occasional extreme values in the unexplained component due to the large kurtosis commonly observed in financial return data (Mandelbrot, 1963).

Herein, we proceed with the above parametrization.  Otherwise, we note that the Particle Learning work of Carvalho, Johannes, Lopes, and Polson (2010) could be further applied to the DPM estimation algorithm to allow estimation of the error parametrization at the same time as the estimation of the latent weight values.

In the following sections, our methods are used to estimate weights on the set of palette assets for the aggregate hedge fund return index.  Then, using these weights, the model is used to forecast the one-step-ahead predicted weights to create a time series of ``forecasted'' index returns.  These returns are plotted along with the observed index returns to compare how precise the estimation is in terms of forecasted return accuracy.

\subsection{Rolling CLS \& ICLS}

\begin{figure}[h]
\centering
\includegraphics[width=0.7\textwidth]{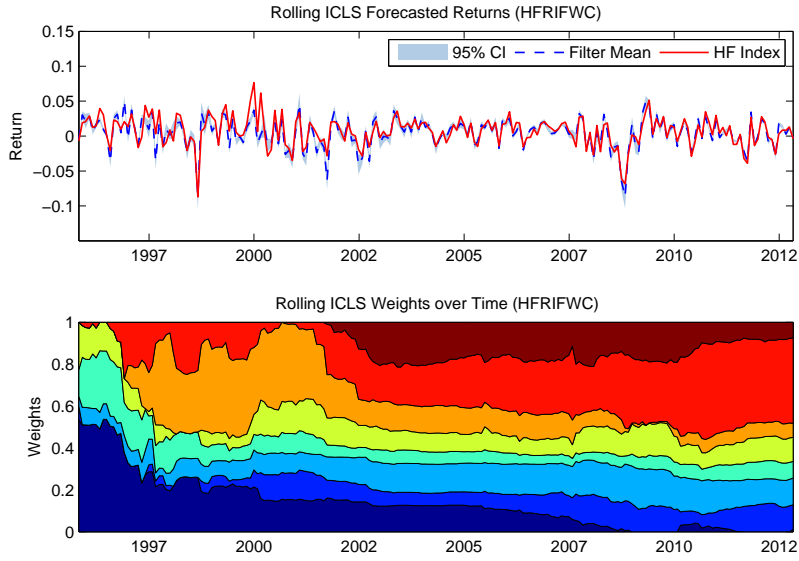}
\caption{\textsl{\small{Forecasted Returns \& Estimated Weights - Rolling ICLS (HFRIFWC)}}}
\label{1_comboICLS_HFRIFWC}
\end{figure}

From the previous sections, the ICLS has strictly produced more accurate results than the CLS without the positivity constraint.  Hence for conciseness, only the ICLS results are exhibited.  Figure \ref{1_comboICLS_HFRIFWC} shows the forecasted returns and estimated weight plots for the Hedge Fund Research Fund Weighted Composite Index (HFRIFWC) using the rolling ICLS estimation method.

Notably, excessively large weights are placed on high yield bonds and short-term treasury securities.  As well, we observe occasional periods with very large jumps in the portfolio weights.  This effect is caused by multicollinearity between the palette asset returns, and therefore the static OLS procedure has a difficult time separating the ultimate return contribution of specific assets.  Because of this, the resulting forecasted returns do no track the index well, thereby inferring poor asset class weight estimates.

Nevertheless, we see a large increase in the weight on short-term treasury bills around the recent economic downturn.  High yield bonds have a very large weight until 2002.  Equity investments seem to be originally focused in US stocks, with a general transition to emerging markets and Europe, Australasia, and the Far East (EAFE) investments over the sample period.  Finally, the plot shows an extra large investment weight in municipal bonds, with a large spike starting around 2001.

This is used as a baseline to see how the three dynamic models compare to this static model's estimation approach.

\subsection{Constrained Kalman Filter}

\begin{figure}[h]
\centering
\includegraphics[width=0.7\textwidth]{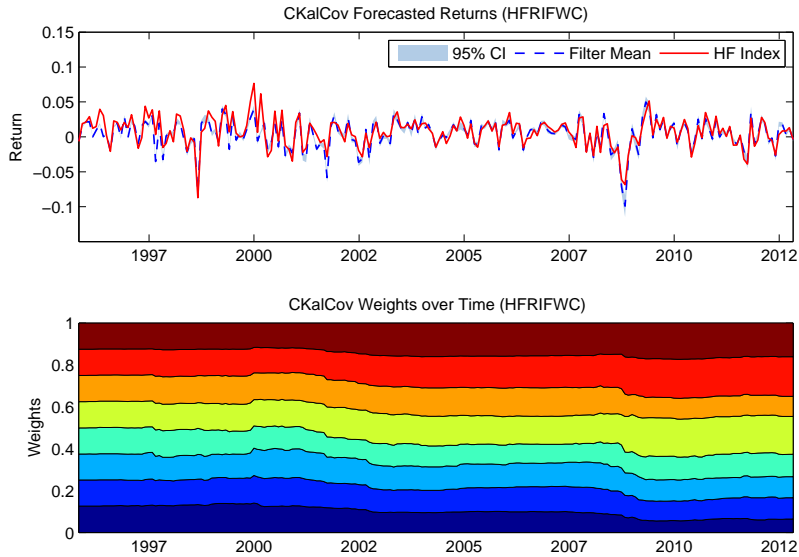}
\caption{\textsl{\small{Forecasted Returns \& Estimated Weights - CKalCov (HFRIFWC)}}}
\label{1_comboKalman_HFRIFWC}
\end{figure}

Since the results for the CKalCov and CKalProj methods are very similar, we only exhibit CKalCov here.  Figure \ref{1_comboKalman_HFRIFWC} shows the forecasted returns and estimated weight plots for the Hedge Fund Research Fund Weighted Composite Index (HFRIFWC) using the CKalCov estimation method.

The Constrained Kalman Filter immediately looks like a superior model from the forecasted return comparison perspective.  There are periods of time where it is inaccurate, but it is much better than the rolling OLS based technique.  However, the estimated weights stay alarmingly even over the entire time horizon.  This is caused by the spherical and time-invariant weight transition covariance structure combined with the contemporaneous correlation observed across financial asset returns.  There are some values that increase or decrease, and not surprisingly, these are generally consistent with the direction of change in the weights from the rolling regressions.  That is, there is larger weight placed on municipal bonds and short term treasuries, while a slowly decreasing weight is placed on US equities.

The most noteworthy attribute of the estimation is the small, but clear jumps in the weights around the dates of economic downturns, thereby indicating a shift in asset allocation occurring in the hedge fund industry in those periods.

\subsection{Dirichlet Portfolio Model}

\begin{figure}[h]
\centering
\includegraphics[width=0.7\textwidth]{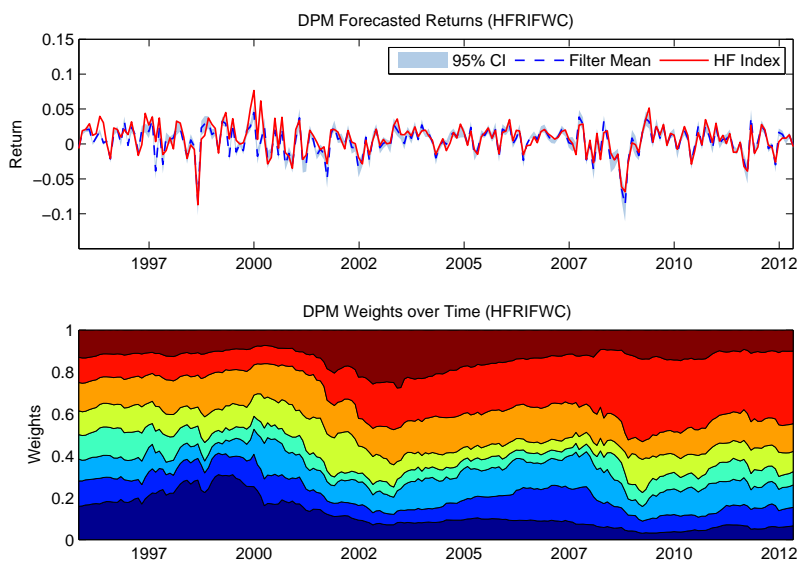}
\caption{\textsl{\small{Forecasted Returns \& Estimated Weights - DPM (HFRIFWC)}}}
\label{1_comboDPM_HFRIFWC}
\end{figure}

Figure \ref{1_comboDPM_HFRIFWC} shows the forecasted returns and estimated weight plots for the Hedge Fund Research Fund Weighted Composite Index (HFRIFWC) using the DPM estimation method.

Looking at the forecasted returns plot, the DPM does an excellent job of tracking the forecasted index returns.  In general, we see fewer periods of poor estimation, as we did with the rolling regressions and the Kalman Filter methods.

As well, the weight estimation results are much more dynamic.  Not only does the plot show clear jumps in the weights around economic downturns and also subsequent shifts to short-term treasuries, but also increases in municipal bond investments over the following years.  Again, there is a noticeable shift from US stocks to increasing investments in emerging markets and EAFE seen over the recent years.  Furthermore, a sizable investment in high yield corporate bonds is observed until 2000, when this weight shifted to investment grade corporate credit from 2000-2006, after the tech bubble.Finally, overall equity and fixed income exposures vary with economic cycles, which is consistent with beliefs about widespread portfolio allocation dynamics.

\subsection{Conditionally Normal Dirichlet Portfolio Model}

\begin{figure}[h]
\centering
\includegraphics[width=0.7\textwidth]{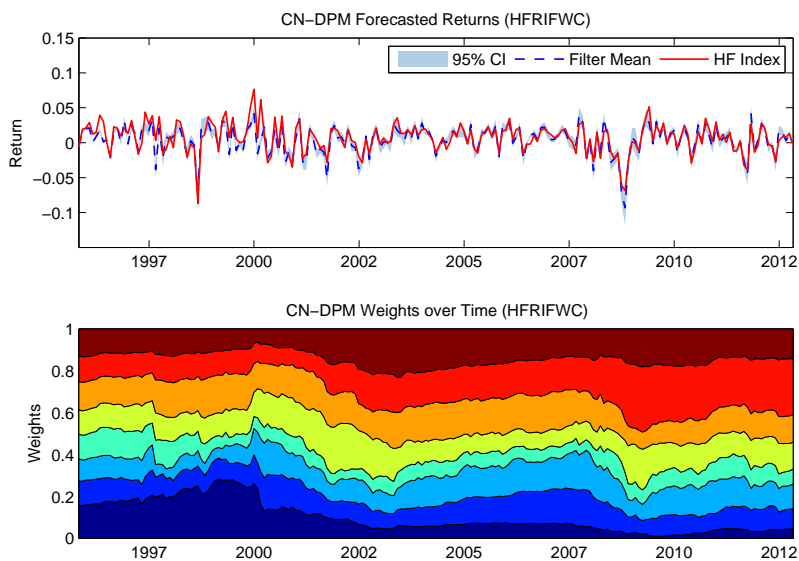}
\caption{\textsl{\small{Forecasted Returns \& Estimated Weights - CN-DPM (HFRIFWC)}}}
\label{1_comboCNDPM_HFRIFWC}
\end{figure}

Figure \ref{1_comboCNDPM_HFRIFWC} shows the forecasted returns and estimated weight plots for the Hedge Fund Research Fund Weighted Composite Index (HFRIFWC) using the CN-DPM estimation method.

The conditionally normal approximation of the DPM also provides accurate forecasted tracking of the hedge fund index.  Generally the same changes in investment patterns are observed.  However, weights remain more evenly distributed across all the asset classes than observed in the original DPM estimates.  This effect is the result of the beta distribution's mode being closer to the bounds of its support.  In other words, its skewed mode pulls values closer to either very large or very small values, whereas a normal approximation distributes probability over the support with no skew.  Lastly, note that this effect does not necessarily imply less or more accurate weight estimation since the forecasted return MAE value is very close to that found with the DPM estimation.

\subsection{Comparison \& Results}

In the previous section, the DPM and CN-DPM demonstrated a superior job in constructing forecasted returns which were consistent with the hedge fund return index observed returns.  Since the true invested hedge fund weights cannot be observed, the accuracy of the weight estimation cannot be directly assessed.  Nevertheless, these forecasted returns give the best proxy for measuring accuracy using the information in this setup.

The DPM and CN-DPM produce asset weight shifts which are generally consistent the broad beliefs about how investment managers have shifted around their asset allocation weights over the last 15 year period.  Examining the estimated changes in weights, perhaps the most interesting pattern is the systematic increase in investment in municipal bonds during recessions, and then a subsequent decrease during economic recovery.  This is consistent with the perceived view that municipal bonds are a relatively safe investment, and therefore during times of high economic uncertainty they provide a reasonably safe investment vehicle.  This is supported by Appleson, Parsons, and Haughwout (2012) who find that municipal defaults are less likely connected to economic downturns than defaults on corporate bonds.  Therefore, observing hedge fund managers increasing investment flows to municipal bonds during these periods is easily rationalized.  Nevertheless, the low risk nature of municipal bonds has come under much debate, especially since late 2007 to early 2008 when many municipal bond prices declined without seeing relative increases on similar duration swap contracts used to hedge interest rate risk (Deng and McCann, 2012).

Furthermore, a comparison of the sample statistics for the estimation methods is shown in Table \ref{1_bigComp}.  This table provides a comparison of the MAE, RMSE, $R^2$, correlation, mean, and standard deviation of the forecasted returns of the six estimation methods.  Note that non-forecasted returns are constructed by applying the estimated weights to the same period returns, while the forecasted returns are the forecasted weights applied to the one-period-ahead returns.  Here, the DPM and CN-DPM consistently outperform the other methods in terms of tracking error, mean absolute error, correlation, and $R^2$.  As well, they generally do the best in replicating the respective mean and standard deviation values.

In all of the above estimation techniques, we point out that large weight is estimated on short-term treasury securities.  As identified in Getmansky, Lo, and Makarov (2004), hedge fund managers commonly employ smoothing in their monthly self reported returns.  Since these treasury assets are perceived to be risk free and have a very low return volatility as compared to the remaining assets, artificially low volatility in the hedge fund index can lead to a larger weight estimated on assets with low volatility themselves, like the short-term treasuries.  From the hedge fund managers' perspective, this smoothing has the effect of improving their funds' observed risk-adjusted performance.  Second, this self reported return smoothing can arise from the pricing of illiquid assets (Fisher et al., 2003; Kadlec and Patterson, 1999).  Therefore, when using self reported hedge fund return data, it is common to estimate a desmoothing model on the return data.  When implementing the model from Getmansky, Lo, and Makarov (2004) on the hedge fund return index, the resulting portfolio weight estimate on the short-term treasuries decreases significantly, while the weights on the remaining assets scale up, proportionately to each other.

\begin{table}[!ht]
\centering
\footnotesize
\begin{tabular}{lccccccc}
&\multicolumn{7}{c}{\textbf{HFRIFWC}}\\
\cline{2-8}
\textbf{Non-Forecasted}&\textbf{Index}&\textbf{DPM}&\textbf{CN-DPM}&\textbf{CLS}&\textbf{ICLS}&\textbf{CKalCov}&\textbf{CKalProj}\\
\hline
Mean&0.00748&0.00457&0.00460&0.00516&0.00478&0.00430&0.00431\\
Standard Deviation&0.02100&0.02014&0.02003&0.02128&0.02100&0.02068&0.02078\\
RMSE&0&0.00921&0.00917&0.00970&0.00980&0.01021&0.01027\\
Mean Abs Error&0&0.00626&0.00644&0.00699&0.00670&0.00752&0.00756\\
Correlation&1&0.90893&0.90949&0.8993&0.89765&0.88967&0.88875\\
$R^2$&1&0.81062&0.81240&0.78994&0.78569&0.76718&0.76474\\
\\
&\multicolumn{7}{c}{\textbf{HFRIFWC}}\\
\cline{2-8}
\textbf{Forecasted}&\textbf{Index}&\textbf{DPM}&\textbf{CN-DPM}&\textbf{CLS}&\textbf{ICLS}&\textbf{CKalCov}&\textbf{CKalProj}\\
\hline
Mean&0.00748&0.00414&0.00408&0.00477&0.00447&0.00412&0.00415\\
Standard Deviation&0.02100&0.02079&0.02084&0.02265&0.02181&0.02116&0.02125\\
RMSE&0&0.01049&0.01074&0.01181&0.01145&0.01076&0.01079\\
Mean Abs Error&0&0.00737&0.00760&0.00844&0.00806&0.00790&0.00791\\
Correlation&1&0.88439&0.87900&0.86160&0.86529&0.87986&0.87969\\
$R^2$&1&0.75509&0.74326&0.68966&0.70851&0.74224&0.74125\\
\end{tabular}
\normalsize
\caption{\textsl{\small{Replication Summary Statistics (Monthly)}}}
\label{1_bigComp}
\end{table}

\subsection{Negative Portfolio Weights}

Throughout this paper, we have assumed and motivated the restriction of non-negativity on the latent asset class weights.  Nevertheless, it can be mentioned that an estimation method which allows for negative weights can easily be constructed in the style of the DPM setup.  The idea is to construct two portfolios, one for long (positive) positions, and another for short (negative) positions.  This allows the long portfolio to capture the variation in the hedge fund index explainable by the positive asset returns, as well as the short portfolio to capture the variation explainable by the negative asset returns.  Then, with estimated distributions for these sets of positive and negative weights, we can estimate a time varying combination factor used to obtain an overall portfolio weighting, thereby potentially increasing the overall explanatory power of the portfolio.

One way to do this is to estimate these separate long/short portfolios in each time period, with respective weights $w_t^+$ and $w_t^-$.  A combined portfolio then can be constructed via $w_t=\left(1+\gamma_t\right)w_t^+-\gamma_tw_t^-$ where the time varying combination factor $\gamma_t$ follows a Gaussian random walk model $\gamma_t\sim N\left(\gamma_{t-1},\sigma_\gamma^2\right)$.  This combination factor can also be estimated via a similar sequential Monte Carlo procedure.  In our estimation problem, not surprisingly, this combination factor was generally found to be $\gamma_t\approx 1$, implying that the aggregate hedge fund industry portfolio does not have negative exposures to these asset classes.

\section{Applications}

\subsection{Replication of Investment Strategy}

The idea of replicating hedge fund investment strategies through low cost, liquid investments is not a new idea.  Many large securities firms currently have products which seek to do exactly this.  Many of these take a bottom-up approach which attempts to identify the types of trades and systematic patterns that funds employ to create their asset allocation, then implement these ideas in an algorithmic manner.  We instead take the top-down approach which is much more statistically sound, as it is attempting to identify component exposures to candidate sets of asset classes, in order to best track the time series of returns.

In order to create a replicating portfolio in this manner, one simply needs a set of relative weights $w^{Rep}$ on the asset set of interest.  Ideally, these weights would be the same as the true hedge fund invested weights, but these are latent.  Therefore, the DPM's expectation of the weights given all observable information up to that time can be used:
\begin{align}
w_{t}^{Rep}&\equiv E[w_t|\mathcal{F}_{t-1}]\nonumber\\
&=\frac{w_{t-1}^{Rep}\circ(1+r_{PA,t-1})}{\sum_{i=1}^nw_{t-1,i}^{Rep}(1+r_{PA,t-1,i})}\nonumber\\
&=\frac{E[w_{t-1}|\mathcal{F}_{t-2}]\circ(1+r_{PA,t-1})}{\sum_{i=1}^nE[w_{t-1,i}|\mathcal{F}_{t-2}](1+r_{PA,t-1,i})}\nonumber\\
&=\dots\nonumber
\end{align}

\begin{figure}[h]
\centering
\includegraphics[width=0.65\textwidth]{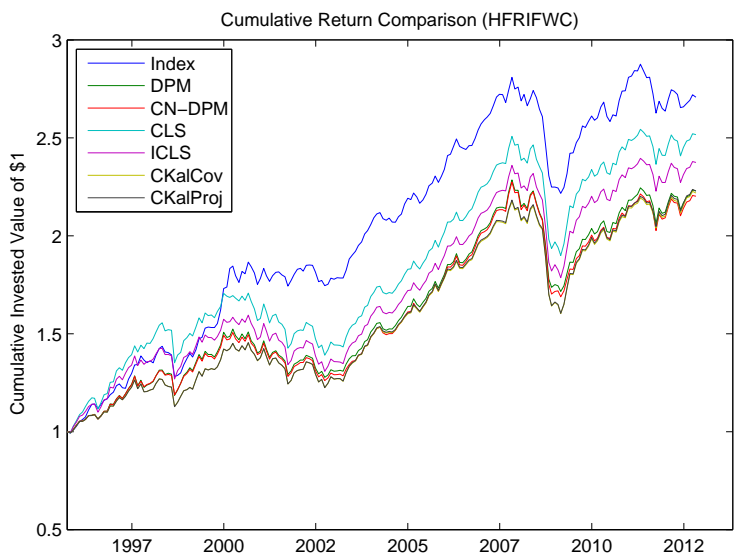}
\caption{\textsl{\small{Cumulative Return Comparison (HFRIFWC)}}}
\label{1_comp_HFRIFWC}
\end{figure}

Using these weights, one can invest in the assets of interest and construct the appropriate replicating portfolios.  Naturally, the goal is to construct portfolios which have very similar returns to investors as the hedge fund indices being replicated.  Since these indices are non-investable, access to these returns is usually obtained through investing in a fund-of-funds, which imposes their own aforementioned layer of fees.  Therefore, the raw index returns are adjusted for these fund-of-fund fees for real comparison purposes.  Nevertheless, we note that there exists upward bias in these index returns that remains unadjusted for.  Figure \ref{1_comp_HFRIFWC} shows a plot of the cumulative return to investors for the adjusted HFRIFWC Index with an initial investment of $\$1$.

The DPM and CN-DPM do a very good job in replicating the return series.  Notably, the rolling CLS does the worst, due to its lack of positivity restriction on the weights, resulting in in-sample over-fitting.  The CKalCov and CKalProj methods do a decent job, however their poor estimation of weights contributes to tracking inaccuracy and under-performance before 2000.

Finally, it is important to note that we do not take a stand on whether implementation of this replication is a good investment strategy.  The answer to this lies in whether the hedge fund industry delivers superior risk adjusted returns.  The answer to that is outside the scope of this paper.

\subsection{Intraperiod Return \& Volatility Approximation}

Since funds only disclose their investment returns at discrete intervals, usually monthly or quarterly, it is difficult for investors to know how their invested capital is performing in the time between.  They could have gained or lost a lot of wealth over the course of a few days, but they will not realize this information for potentially months later.  This presents an informational problem, since the knowledge of this investment performance has clear implications for consumption decisions in the current period.

As well, it is very common to see portfolio managers for large pension funds, endowments, family offices, etc. not only invest in individual assets, but also other investment managers.  Therefore, there is value in knowing how their less liquid and transparent hedge fund investments are performing in order to more appropriately manage risk in the rest of their overall portfolio.  Hence, having an approximation of these intraperiod hedge fund return and volatility values is of great value.

With the DPM setup, we have an effective way of approximating these intraperiod returns.  Consider the following setup where we want to approximate the intraperiod returns for $\tau$ units of time past reporting period time $t$:

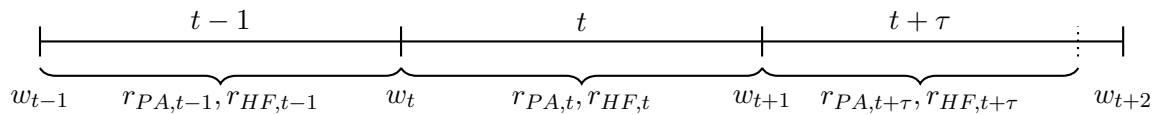
\begin{figure}[H]
\centering
\begin{tikzpicture}[xscale=1.2]
\tikzset{position label/.style={below=10pt,text height=2ex,text depth=1ex}}
\draw [thick] (0,0) -- (12,0);
\draw [thick] (0,-0.2) -- (0,0.2);
\draw [thick] (4,-0.2) -- (4,0.2);
\draw [thick] (8,-0.2) -- (8,0.2);
\draw [dotted,thick] (11.5,-0.2) -- (11.5,0.2);
\draw [thick] (12,-0.2) -- (12,0.2);
\node [position label] (t1) at (0,0) {$w_{t-1}$};
\node [position label] (t2) at (4,0) {$w_{t}$};
\node [position label] (t3) at (8,0) {$w_{t+1}$};
\node [position label] (tt) at (11.5,0) {};
\node [position label] (t4) at (12,0) {$w_{t+2}$};
\draw [thick,decoration={brace,amplitude=6pt,mirror},decorate] (t1.north) -- (t2.north) node[midway,below=5pt] {$r_{PA,t-1},r_{HF,t-1}$};
\draw [thick,decoration={brace,amplitude=6pt,mirror},decorate] (t2.north) -- (t3.north) node[midway,below=5pt] {$r_{PA,t},r_{HF,t}$};
\draw [thick,decoration={brace,amplitude=6pt,mirror},decorate] (t3.north) -- (tt.north) node[midway,below=5pt] {$r_{PA,t+\tau},r_{HF,t+\tau}$};
\node[above] at (2,0) {$t-1$};
\node[above] at (6,0) {$t$};
\node[above] at (9.75,0) {$t+\tau$};
\end{tikzpicture}
\caption{\textsl{\small{Intraperiod Timeline Illustration}}}
\label{1_intraperiod}
\end{figure}

We are interested in determining the value of the holding period return and volatility over time period $t+\tau$ given the observed information at time $t+\tau$.  That is, we want to estimate $E[r_{HF,t+\tau}|\mathcal{F}_{t+\tau}]$ and $Vol[r_{HF,t+\tau}|\mathcal{F}_{t+\tau}]$.  However, note that $\mathcal{F}_{t+\tau}=\left\{\mathcal{F}_{t},r_{PA,t+\tau}\right\}$ since the only new information observed since time $t$ is the return on the palette assets.  Since the true weights cannot be observed to use in this calculation, we can use our estimated weights from time $t$.  Therefore, the estimate of intraperiod return, using the DPM model's estimates, simply becomes:
\begin{align}
E\left[r_{HF,t+\tau}|\mathcal{F}_{t+\tau}\right]&=E\left[r_{HF,t+\tau}|\mathcal{F}_t,r_{PA,t+\tau}\right]\nonumber\\
&=E\left[w_t|\mathcal{F}_{t}\right]'r_{PA,t+\tau}\nonumber
\end{align}

Similarly, an estimate of the intraperiod volatility can be computed:
\begin{align}
Vol\left[r_{HF,t+\tau}|\mathcal{F}_{t+\tau}\right]&=Vol\left[r_{HF,t+\tau}|\mathcal{F}_t,r_{PA,t+\tau}\right]\nonumber\\
&=\sqrt{E\left[w_t|\mathcal{F}_{t}\right]'\hat{\Sigma}_{r_{PA,t+\tau}} E\left[w_t|\mathcal{F}_{t}\right]+\sigma^2_\epsilon}\nonumber
\end{align}
where $\hat{\Sigma}_{r_{PA,t+\tau}}$ is an intraperiod covariance matrix for the palette assets.  This matrix for the latent covariance structure can be constructed in various ways, including but not limited to Stochastic Volatility (Jacquier, Polson, and Rossi, 1994, 2004) and DCC-GARCH (Engle, 2002) approaches.  Furthermore, the same expression above can be used to compute \emph{forecasted} hedge fund volatility if the current period is taken to be time $t$ and we want to forecast $\tau$ time into the future.

\section{Conclusion \& Discussion}

This paper has presented a Bayesian dynamic model, the Dirichlet Portfolio Model (DPM), for the hedge fund industry weight transition process and aggregate return observations.  We then exhibited a numerical solution to this model using sequential Monte Carlo methods, as well as a conditionally normal approximation (CN-DPM) which was solved analytically.  In order to motivate the appropriateness of this dynamic model, other models and their respective solutions were compared.  The simulated and model hedge fund results showed that both with simulated or real assets, as well as under simulated or model trading, the DPM produces more accurate estimates of the underlying weights as compared to the results produced by the other estimation methods.  Overall, the DPM provided superior results across various measures of suitability.  Interestingly, the estimation results on the the hedge fund industry aggregate return index identify a systematic increase in exposure to municipal bonds during economic downturns, and a subsequent decrease in exposure during economic recovery periods, which is consistent with the notion that defaults on municipal bonds are less connected to economic downturns than defaults on corporate bonds.

From the foundational DPM, there are many future extensions from this starting point.  One of the challenges of the DPM estimation procedure is having to pre-specify the distributional error parameters, $\alpha$, $\sigma_\epsilon^2$, and $\nu$.  Just as there is value in obtaining the latent weight estimates, it would also be insightful to learn the magnitude of these tuning parameters during the estimation process.  This can be achieved by applying the parameter learning concepts from Carvalho, Johannes, Lopes, and Polson (2010), Storvik (2002), Fearnhead (2002), or Liu and West (2001).

Furthermore, we identify applications of this methodology to both creation of hedge fund industry replicating portfolios, as well as intra-reporting-period return and volatility estimation.  There is large value in being able to approximate these intra-reporting-period returns for both current consumption choices and various risk management decisions.  As well, being able to create replicating portfolios from the asset class decomposition has the potential to construct more transparent portfolios with much lower cost structures.  Therefore, the Dirichlet Portfolio Model is a convenient technique for decomposing unobservable portfolio compositions, allowing for future analysis on the dynamics of these weight processes.  

\section{References}

\small
\begingroup
\renewcommand{\section}[2]{}

\endgroup

\end{document}